\documentclass[aps,prl,twocolumn,floatfix]{revtex4-2}
\usepackage[utf8]{inputenc}
\usepackage[T1]{fontenc}
\usepackage[english]{babel}
\usepackage{mathtools}
\usepackage{newtxtext}  
\usepackage{amsmath}
\usepackage{amsfonts}
\usepackage[varbb,varg]{newtxmath} 
\usepackage{bm,dsfont,eucal}
\AtBeginDocument{\renewcommand{\vec}[1]{\bm{#1}}}

\usepackage[dvipsnames,svgnames,x11names,hyperref]{xcolor}
\usepackage{hyperref}
\hypersetup{ 
    colorlinks=true, linkcolor=NavyBlue, urlcolor=NavyBlue, citecolor=NavyBlue
}

\usepackage{physics,braket,siunitx,slashed}
\usepackage[capitalise]{cleveref}

\begin{document}

\title{Momentum-Space AC Josephson Effect and Intervalley Coherence in Multilayer Graphene} 

\author{Mainak Das and Chunli Huang}
\affiliation{Department of Physics and Astronomy, University of Kentucky, Lexington, Kentucky 40506-0055, USA}

\date{\today} 

\begin{abstract}
Electron transport driven by the phase coherence and interference of quantum many-body wavefunctions is a fascinating phenomenon with potential technological significance. Superconductivity, for example, enables dissipationless transport through macroscopic phase twisting. Similarly, in charge-density waves, once the phase degree of freedom—representing the collective position of electrons relative to the lattice—is depinned, it generates characteristic broadband noise and intriguing AC-DC interference patterns. In this work, we point out a phase-coherent dynamics in the  intervalley coherent (IVC) state, also known as the bond-ordered or Kekul\'e distorted state, frequently reported in rhombohedral multilayer graphene.  Under a static magnetic field, the IVC state responds with an oscillating intervalley current, which in turn causes oscillating orbital magnetization, thereby inducing a detectable AC Hall effect. This mechanism mirrors the AC Josephson effect observed in superconductors but now happening in momentum space. In this analogy, the static magnetic field acts as the DC voltage, while the oscillating intervalley current assumes the role of the AC Josephson current. We present detailed microscopic calculations for all the parameters of the phase-number free-energy in rhombohedral trilayer graphene, predicting an orbital magnetization oscillation frequency of approximately 12 GHz at 0.1 Tesla.
We comment on this phase-coherent dynamics in other 2D materials like twisted homobilayer transition metal dichalcogenides.

\end{abstract} 

\maketitle


\textit{Introduction: }Magnetism arising from spin and valley ordering in lightly-doped multilayer graphene systems, both with \cite{PhysRevB.99.201408,nuckolls2024microscopic,liu2021orbital,tschirhart2021imaging,bhowmik2022broken,chen2022tunable,PhysRevLett.124.187601,lin2022spin,wu2020ferromagnetism,wagner2022global,he2021competing,wu2021chern} and without moir\'e potential \cite{zhou_half_2021,arp2024intervalley,zhou_superconductivity_2021,de_la_Barrera_2022,doi:10.1126/science.abm8386,seiler2022quantum,lee2019gate,lee2014competition,liu2023spontaneous,Myhro_2018,wirth2022experimental,PhysRevB.79.125443,Han_2023,han2023orbital,morissette2025intertwinednematicitymultiferroicitynonlinear,PhysRevB.111.075103}, has become a well-established phenomenon following the discovery of the integer quantum anomalous Hall effect in twisted bilayer graphene \cite{sharpe2019emergent,serlin2020intrinsic,tseng2022anomalous,geisenhof2021quantum}. There are three primary types of spin-valley ordered states: spin-polarized ferromagnets, valley-polarized orbital ferromagnets, and the intervalley-coherent (IVC) state. Spin-polarized ferromagnets typically expand their phase boundaries under in-plane magnetic fields, whereas valley-polarized orbital ferromagnets are known for exhibiting a strong anomalous Hall effect. These states, characterized by Ising-like order parameters, also generate magnetic stray fields detectable via nano-SQUID \cite{Wernsdorfer_2009,GRANATA20161,finkler2010self,patterson2024superconductivityspincantingspinorbit}. In contrast, the intervalley-coherent (IVC) state, characterized by its XY-like order parameter, remains elusive and is typically identified by eliminating other symmetry-broken states \cite{wu2021chern,zhou_half_2021,arp2024intervalley} and through comparison with mean-field theory \cite{PhysRevLett.124.166601,PhysRevX.10.031034,huang2023spin,chatterjee_inter-valley_2021,PhysRevB.110.035103}. Theories suggest that IVC fluctuations could be the microscopic origins of superconductivity \cite{vituri2025incommensurate,chatterjee2022inter} in multilayer graphene and homobilayer moir\'e transition metal dichalcogenides \cite{fischer2024theory}.
The IVC state was previously thought to exhibit no prominent transport signatures and can mainly be detected through STM experiments \cite{farahi2023broken,liu2022visualizing,nuckolls2023quantum,kim2023imaging,coissard2022imaging}.
 We note that the difference in the dielectric screening environment can influence the observable magnetic grounstates in STM versus transport experiments~\cite{coissard2022imaging,wei2025landau,xu2024influence}.

In this study, we report a unique transport phenomenon associated with the IVC state: an out-of-plane static magnetic field induces an oscillating orbital magnetization, leading to an AC Hall effect, as shown in Fig.~\ref{fig:schematic}. This effect, analogous to AC Josephson effect in a superconductor, can be detected using probes sensitive to finite-frequency Hall conductivity $\sigma_{xy}(\omega)$, such as magneto-optical Kerr effect measurements, and may serve as a definitive diagnostic tool for identifying the IVC state.

\begin{figure}[t]
    \centering
\includegraphics[width=0.9\linewidth]{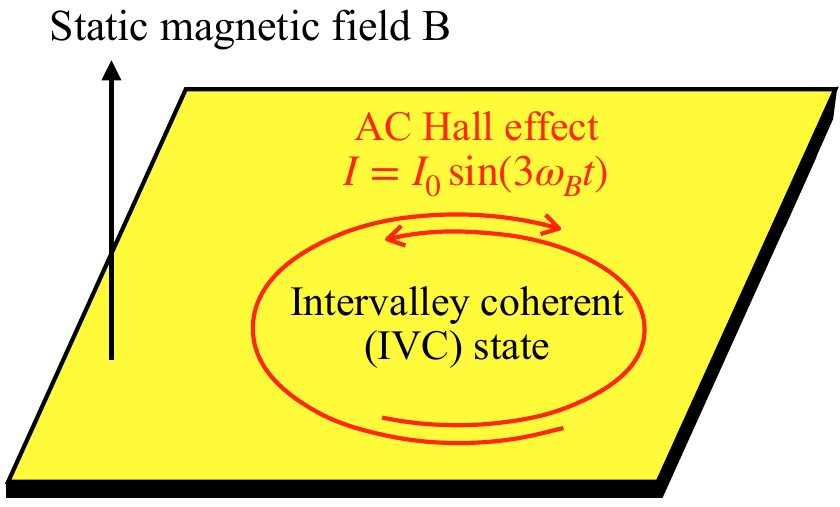}
\caption{When an intervalley coherent (IVC) state is subjected to a static magnetic field, it generates an oscillating orbital magnetization, resulting in a AC Hall current (red arrows) oscillating at frequency $3\omega_B=3g_v\mu_BB/\hbar$,  detectable via probes sensitive to AC Hall conductivity.
}
\label{fig:schematic}
\end{figure}

\textit{Momentum-Space AC Josephson Effect:} 
In multilayer graphene with a low electron density of about $10^{-4}$ electrons per carbon atom (typical in most experiments),  electrons occupied a tiny fraction of the Brillouin zone, specifically around the opposite corners, which are denoted  as valleys $\tau={K, K'}$.
To make the connection to the BCS wavefunction more transparent, we now perform a particle-hole transformation in one of the valleys and denote its transformed vacuum by $|\Psi_{vac}\rangle$. Then, the IVC groundstate can be written analogously to a BCS superconductor, except that the pairing occurs between electrons and holes in opposite valleys, rather than between electrons with opposite spins and momenta:
\begin{equation} \label{eq:exciton_wf}
    |\Psi_\phi\rangle = \prod_{\vec{k}}(u_{\vec{k}} + v_{\vec{k}}e^{i\phi} c_{\vec{K}+\vec{k}}^\dagger c_{-\vec{K}+\vec{k}})|\Psi_{vac}\rangle.
\end{equation}
Here, $\vec{k}$ runs over a small subset of wave-vectors (reflecting the low electron density) and $u_{\vec{k}}=\cos\theta_{\vec{k}}/2$ and $v_{\vec{k}}=\sin\theta_{\vec{k}}/2$ are real coefficients representing valley-mixing angle.  The operator $c_{\vec{K}+\vec{k}}^\dagger c_{-\vec{K}+\vec{k}}$ acting on $|\Psi_{vac}\rangle$ creates an exciton with wave-vector $2\vec{K}$. Thus, $|\Psi_\phi\rangle$  describes a condensate of such $2\vec{K}$ excitons with $\phi$ representing the global phase of the condensate. The spin degree of freedom is not important to our discussion, and it suffices to treat the electrons and holes as having the same spin projection. 
Analogous to the BCS state in superconductors, the state $|\Psi_\phi\rangle$ possesses a well-defined global phase but an indefinite particle number. The wavefunction with a definite number of excitons can be obtained by integrating $|\Psi_{\phi}\rangle$ over the global phase after multiplying by an appropriate phase factor, effectively "localizing" the state in particle-number space \cite{anderson1966considerations}: $ \ket{\Psi_{N_{\tau}}} = \int  d\phi \, c_{N_\tau}\, e^{iN_{\tau}\phi}\ket{\Psi_{\phi}}$. This exciton picture of electron–hole pairing at distinct wave-vectors has been widely used to describe one-dimensional charge-density-wave (CDW) metals like NbSe$_3$ (e.g., Ref.~\cite{RevModPhys.60.1129}), where each exciton carries wavevector $2k_F$, with $k_F$ being the Fermi wavevector. However, there is an important difference between a 1D CDW and our $2\vec{K}$ exciton condensate in higher dimensions.
In 1D, the number eigenstate with $N$ excitons carries a net velocity of $2N\hbar k_F/m^*$ ($m^*$ is the effective mass) and thus describe a current carrying state. In our case, however, a state with $N_\tau$ excitons $|\Psi_{N_{\tau}}\rangle$ does \emph{not} carry a net velocity or current if electrons in each valley remain in independent chemical equilibrium. The reason is that, in higher dimensions, the total velocity in each valley vanishes upon integrating over the Fermi surface in chemical equilibrium,
 $ \int d^2k (\partial \epsilon_{k,\tau}/\partial k) n_F(\epsilon_{k,\tau}-\mu)=0$. Therefore, although the exciton number eigenstate possesses finite crystal momentum ($\bra{\Psi_{N_{\tau}}}\vec{k}\ket{\Psi_{N_{\tau}}}\neq0$) it generally does not produce a measurable net current. However, multilayer graphene is an exception. Here, Bloch states at $\pm \vec{K}$ points
can acquire opposite \emph{orbital angular momentum} \cite{xiao2007valley} if the Dirac points are gapped by a displacement field, interaction effects, or both. This means that the exciton number eigenstate $|\Psi_{N_{\tau}}\rangle$  carries a finite orbital moment and thus has measurable consequences. This property of opposite orbital angular momenta in different valleys is also possible in other 2D materials like twisted MoTe$_2$
and twisted WSe$_2$  when their valley-projected Hamiltonians host moir\'e minibands with finite Chern numbers \cite{reddy2023fractional,wu2018hubbard}. Because electrons and holes possess opposite orbital moments, an external magnetic field 
$B$ shifts their energy levels in opposite directions, effectively acting like a valley-contrasting chemical potential \cite{xiao2007valley}:
\begin{equation} \label{eq:OM}
    F_B =  - MB\equiv -g_v\mu_B N_{\tau} B
\end{equation}
where $M=g_v\mu_B N_{\tau}$ is the net orbital moment. We will substantiate this relationship through a detailed microscopic calculation in the following section.


Another unusual aspect of a $2\vec{K}$ exciton condensate is the presence of intrinsic intervalley tunneling from Umklapp scattering. While $2\vec{K}$ and $4\vec{K}$ are not reciprocal lattice vectors, $6\vec{K} = 4\vec{b}_1 +2 \vec{b}_2$ \emph{is}, where $\vec{b}_{1,2}$ are the two primitive reciprocal lattice vectors \cite{castro2009electronic}. Consequently, lattice translation symmetry prohibits two-body Umklapp scattering \cite{levitov2016electron} but allows three-body Umklapp scattering: $\hat{H}_{\text{ump}}\sim c_K^\dagger c_K^\dagger c_K^\dagger  c_{K'}  c_{K'}  c_{K'} + h.c. $. This process is analogous to the Josephson coupling in Josephson junctions, which maintains phase coherence across a junction by tunneling Cooper pairs. Similarly, three-body Umklapp scattering enforces phase coherence in the IVC exciton condensate by tunneling three electrons from one valley to another. This interaction introduces an intrinsic potential energy that maintains phase coherence:
\begin{equation} \label{eq:F_ump}
    F_{\text{ump}} = -\gamma \cos(3\phi)
\end{equation}
where $\gamma$ represents the strength of the 3-body scattering, to be estimated in the following section.


%
\begin{figure}
    \centering
    \includegraphics[width=0.8\linewidth]{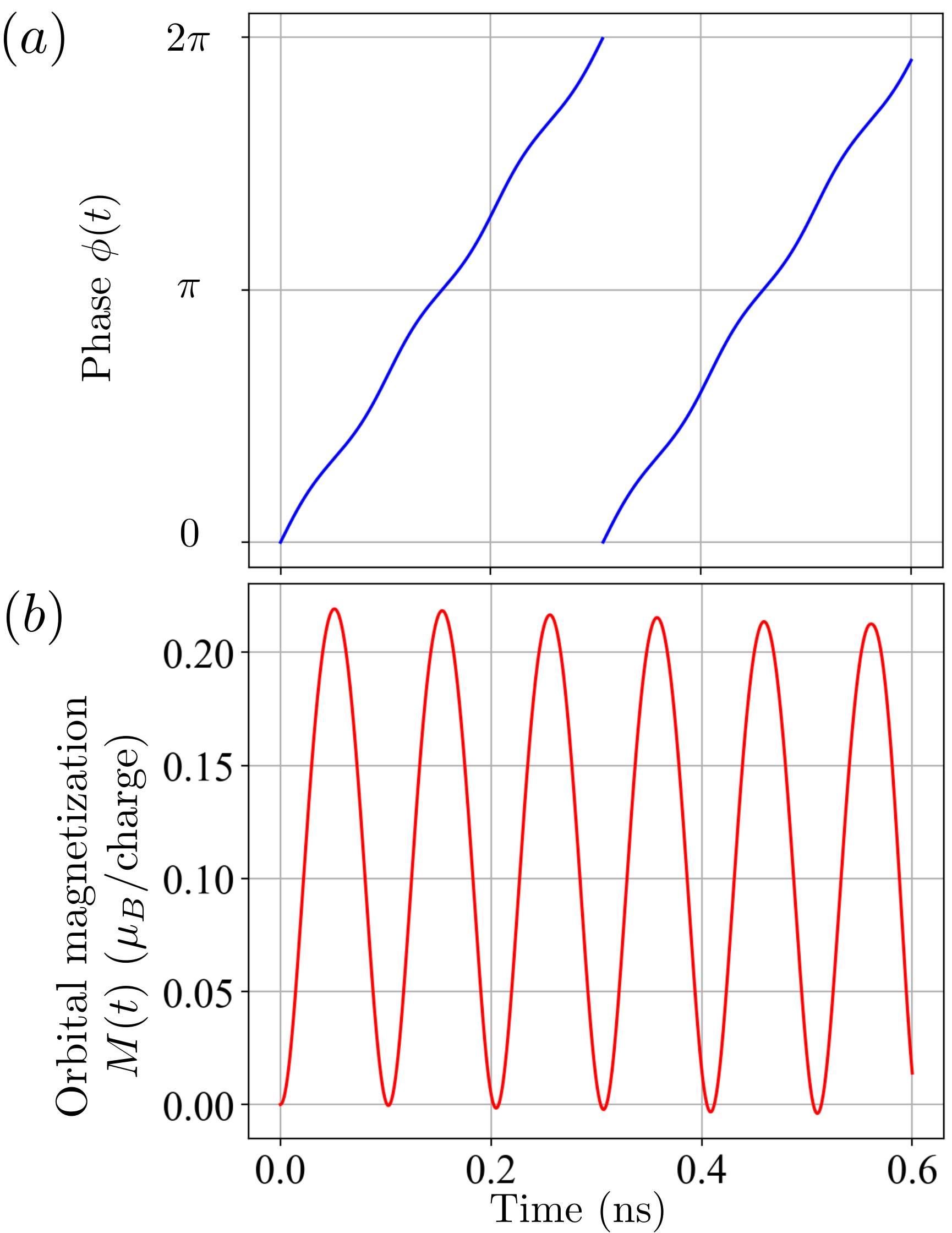}
    \caption{Uniform solution of Eq.~\eqref{eq:Jospephson_1} and \eqref{eq:Jospephson_2} with initial conditions $\phi(0)=N_\tau(0)=0$. Here $M(t)=g_v\mu_B N_\tau(t)$ is plotted for IVC  in rhombohedral trilayer graphene with $\chi^{-1}=0.1$ meV, $g_v=3$, $\gamma=0.5~\mu$eV, $B=0.1$T and initial conditions $\phi(0)=N_\tau(0)=0$. The amplitude (frequency) of $M$ oscillations  decreases (increases) linearly with $B$.}
    \label{fig:phase_OM_evolution}
\end{figure}
Eq.~\eqref{eq:OM} and \eqref{eq:F_ump}, along with the  superfluid kinetic terms,
\begin{equation} \label{eq:F_0}
    F_0 = \frac{\rho}{2}(\nabla \phi)^2 + \frac{1}{2\chi}N_{\tau}^2,
\end{equation} 
where $\rho$ and $\chi$ are the phase-stiffness constant and susceptibility to create electron-hole imbalance,
form the basis of our analysis on unusual transport phenomena in IVC states. They define a phase-number action, $F_{\text{IVC}}$, that applies to any $2\vec{K}$ exciton-condensate with electrons and holes carrying opposite orbital angular momentum:
\begin{equation}
    F_{\text{IVC}}  =F_0 + F_{\text{ump}} + F_B.
\end{equation}
The phase coherent transport of superfluid can be obtained by treating $N_\tau$ and $\phi$ as dynamical conjugate variables \cite{anderson1966considerations}:
\begin{align} \label{eq:Jospephson_1}
    \hbar \frac{d\phi}{dt}&=-\frac{\partial F_{\text{IVC}}}{\partial N_{\tau}} = -\frac{N_{\tau}}{\chi} + g_v\mu_BB,  \\
     \hbar \frac{dN_{\tau}}{dt}&=\frac{\partial F_{\text{IVC}}}{\partial \phi}= -\rho \nabla^2\phi+ 3\gamma\sin(3\phi).\label{eq:Jospephson_2}
\end{align}

Eq.~\eqref{eq:Jospephson_1} and \eqref{eq:Jospephson_2} represent the main results of our article, describing phase-coherent dynamics in IVC. We focus on a uniform solution with the initial conditions  $\phi(0)=N_\tau(0)=0$. As shown in Fig.~\ref{fig:phase_OM_evolution}, static magnetic field acts like to a chemical potential for the exciton, inducing a nearly linear, coherent phase rotation of the condensate over time. This phase dynamics then subsequently generates an oscillating exciton number $N_{\tau}(t)$, through intervalley tunneling, leading to an oscillating orbital magnetization, $M(t)=g_v \mu_B N_{\tau}(t)$. Table.~I summarizes the parallels between the AC Josephson effect in a superconductor and the analogous phenomenon in our 
$2\vec{K}$ exciton condensate. For small $N_\tau(t)/\chi$, the solution is straightforward: $\phi(t) =\omega_B t$  and $ M(t) = M_0 [1-\cos(3\omega_B t)]$ where $M_0=g_v(\gamma/\hbar \omega_B)\mu_B$ and $\omega_B=g_v\mu_BB/\hbar$. Note the amplitude (frequency) of $M(t)$ decreases (increases) linearly with external field $B$. The minor nonlinear feature in Fig.~\ref{fig:phase_OM_evolution}$(a)$ results from the finite $N_\tau(t)/\chi$.


\begin{table}
\centering
\begin{tabular}{lcc}
\hline
\textbf{Characteristic} & \textbf{Superconductor} & \textbf{IVC} \\
\hline
Static driving field & Voltage (V) & Magnetic field (B) \\
Phase evolution & $\dot \phi = 2eV/\hbar$ & $\dot \phi = g_v\mu_BB/\hbar$ \\
Phase coherence & Josephson coupling & 3-body Umklapp \\
AC response &  Josephoson current & Intervalley current
\end{tabular}
\caption{Comparison of the AC Josephson effect occurring in real space in superconductors with its analog in momentum space in  IVC states. Oscillating intervalley currents lead to oscillating orbital magnetization.}
\end{table}


\textit{Rhombohedral trilayer graphene (RTG):} We now demonstrate how to compute all the parameters in $F_{\text{IVC}}$ for rhombohedral trilayer graphene using a detailed microscopic $K\cdot p$ model. We primarily focus on RTG as it was the first rhombohedral-stacked graphene in which IVC states were experimentally identified \cite{zhou_half_2021}. IVC states have also been observed in other layered graphene structures \cite{zhang2023enhanced,PhysRevLett.131.146601,hagymasi2022observation}.
We start with the many body Hamiltonian $\hat{H}=\hat{H}_{b}+\hat{V}$, where
\begin{equation}
     \hat{H}_b=\sum_{\vec{k},\tau,s,\sigma}\; h_{\sigma,\sigma'}(\tau k_x,k_y) c^\dagger_{\tau,\sigma,s}(\vec{k})  c_{\tau,\sigma',s}(\vec{k})\\
\end{equation}
encapsulates all the intricacies of the Slonczewski-Weiss-McClure (SWMC) parameters and the important role of the electric-displacement field in enhancing the density of states, see Ref.~\cite{supmat}. This is represented by a 
$24\times24$ matrix, reflecting the six sublattices $\sigma$, two spin $s$, and two valley $\tau$ degrees of freedom. The interaction is modeled as a density-density type:
\begin{equation}\label{eq:Coulomb-pot}
    \hat{V} = \frac{1}{2} \sum_{\vec{q}} V_{\vec{q}}
    \hat{n}_{\vec{q} } \hat{n}_{-\vec{q}}, 
\end{equation}
where $\hat{n}_{\vec{q}} = \sum_{\vec{k},\sigma,\tau,s} c_{\tau,\sigma,s,\vec{k}+\vec{q}}^\dagger  c_{\tau,\sigma,s,\vec{k}} $ and $V_{\vec{q}}=2\pi k_e\tanh{(|\vec{q}|d)}/(\epsilon_r|\vec{q}|)$ is the dual gated screened Coulomb potential. The layer-dependent anisotropy interaction is suppressed in the strong electric-displacement limit and it is not important for discussing IVC transport. Here, $k_e=1.44$eVnm is the Coulomb constant, $\epsilon_r=15$ is the screening constant, and $d=5$ nm is the distance from the gate to the material.
Next, we solve this Hamiltonian using the Hartree-Fock approximation, $\hat{H}\approx \hat{H}_{MF}$ where the matrix elements of $\hat{H}_{MF}$ are determined self-consistently at effectively zero temperature (with negligible temperature for numerical smearing purposes) and at a fixed electron density $n_e$. For detail discussion of the self-consistent Hartree-Fock calculations, including the discussion of Fast Fourier Transformation trick to obtain the self-energy, see Ref.~\cite{wolf2024magnetismdiluteelectrongas}. 
The groundstate thus obtained at $(n_e, U)=-2.7\times 10^{11} \text{cm}^{-2},15$ meV is a spin-polarized intervalley coherent (IVC) state, consistent with previous mean-field calculations \cite{huang2023spin,chatterjee_inter-valley_2021,PhysRevB.110.035103} and in agreement with with experimental \cite{zhou_half_2021,arp2024intervalley} outcomes through a process of eliminating other possibilities. In this calculations, we did not perform a particle-hole conjugation in one of the valleys, following the convention in prior works to allow for direct comparison. However, we note that interpreting the IVC state as an exciton condensate is more natural when such a particle-hole transformation is applied in one of the valleys. A linecut of bandstructure for the IVC running wave is showed in Fig.~\ref{fig:IVC_bandstructure}. 
Next, we introduce a Lagrange multiplier into the Hamiltonian $H$:
\begin{equation}
    \hat{H}_{\lambda} = -\lambda \, (|K\rangle\langle K|-|K'\rangle\langle K'|),
\end{equation} 
and solve it again using mean-field theory to obtain a converged density matrix  $\hat{\rho}_\lambda= \sum_{nk}n_F(\epsilon_{nk}^\lambda)|\psi_{nk}^\lambda\rangle \langle \psi_{nk}^\lambda|$ as a function of $\lambda$. (Here $\epsilon_{nk}^\lambda$ and $| \psi_{nk}^\lambda\rangle $ are the mean-field quasiparticle energy and eigenstate).
The resulting bandstructure with $\lambda=0.05$ is shown in Fig.~\ref{fig:IVC-lagrange-multiplier}$(a)$. 

\begin{figure}
    \centering
    \includegraphics[width=0.8\linewidth]{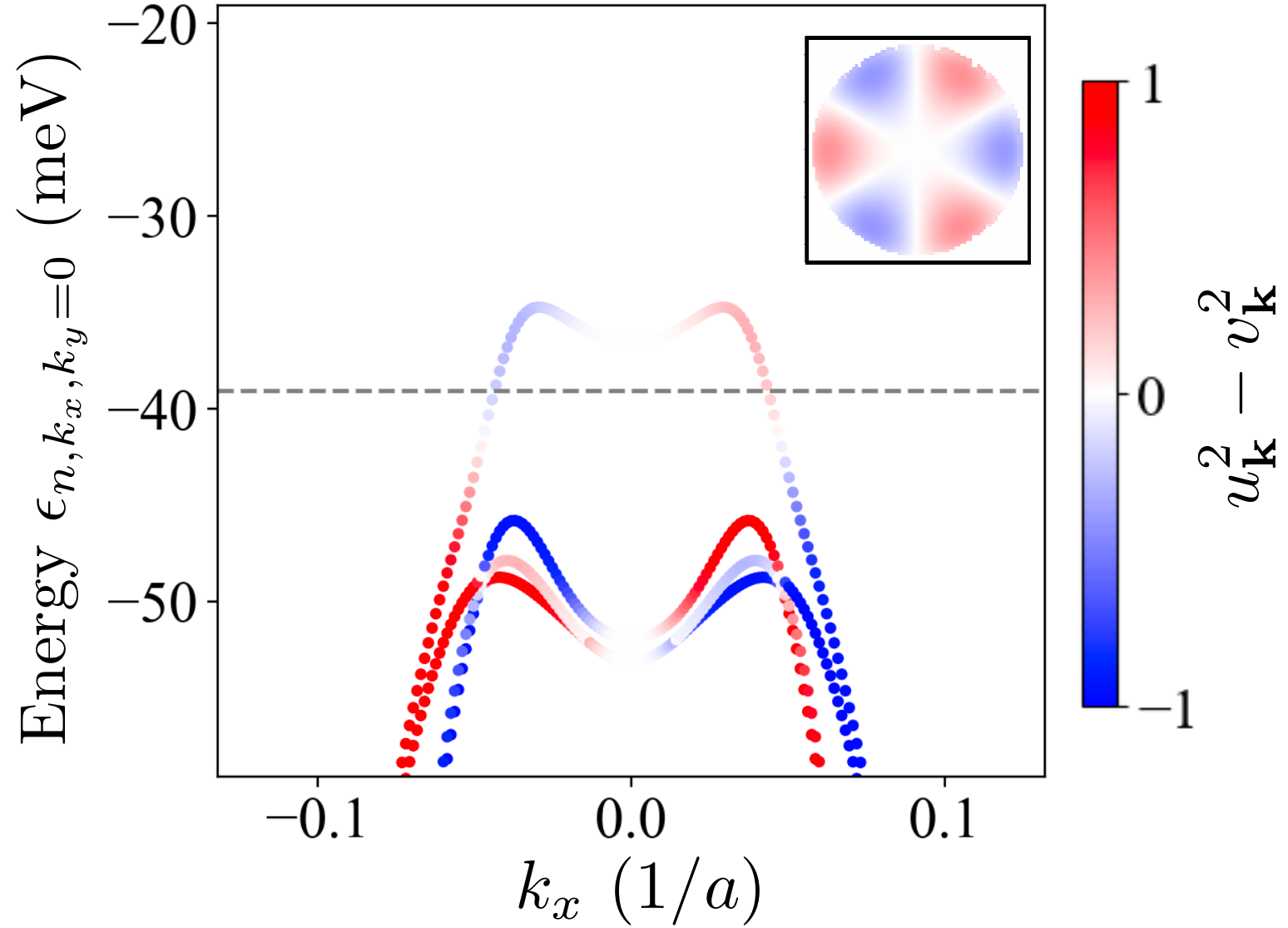}
    \caption{Energy-momentum dispersion of running waves in IVC metal in RTG. The color bar shows cosine of the valley mixing angle, with the inset detailing this angle within the Fermi sea of the top valence band.
    The dashed line marks the Fermi level.
    }
    \label{fig:IVC_bandstructure}
\end{figure}

\begin{figure*}[t]
    \centering   \includegraphics[width=1.00\linewidth]{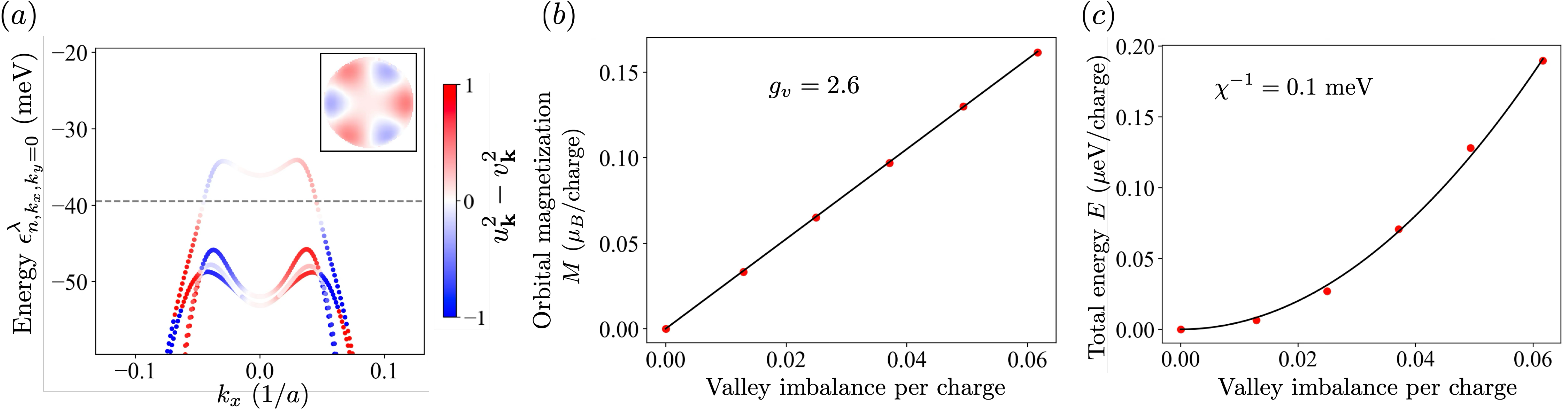}
    \caption{Properties of an IVC state with valley-imbalance $N_\tau$ simulated using a Lagrange multiplier $\lambda$: a) Energy dispersion of an IVC state at $\lambda=0.05$ with an inset showing an increased number valley-imbalance enclosed in the Fermi surface compare to Fig.~\ref{fig:IVC_bandstructure}. Orbital magnetization b) and energy increase c) vs $N_\tau$ at $(n_e, U)=-2.7\times 10^{11} \text{cm}^{-2},15$ meV. $N_\tau$ is a monotonically increasing function of $\lambda$ for small $\lambda$.}
    \label{fig:IVC-lagrange-multiplier}
\end{figure*}

The introduction of $\lambda$ results in a converged solution with a net valley-imbalance, or in exciton language, a net electron-hole imbalance $N_\tau=\text{Tr}[|K\rangle\langle K|-|K'\rangle\langle K'|) \hat{\rho}_\lambda]$. 
 This imbalance causes finite orbital magnetization $M$ due to the opposite orbital moments in opposite valleys, and an energy increase  $E=\frac{1}{2}\text{Tr}[(H_b+H^{MF})\hat{\rho}_\lambda] $ due to the depletion of excitons from the condensate, as shown in Fig.~\ref{fig:IVC-lagrange-multiplier}.
 Importantly, this state we simulated with finite 
$\lambda$ represents a many-body state where the energy increase due to the unpaired excitons of the condensate, $N_\tau^2/(2\chi)$, is counterbalanced by the energy lowering of the orbital magnetization $-MB$.
\footnote{ In valley-pseudospin language, this corresponds to an excited state with a slight valley imbalance, where the valley pseudospin order parameter deviates a little bit away from the XY-plane}. This scenario represents a stationary solution of Eq.~\eqref{eq:Jospephson_1} where $\dot\phi=0$ and $N_\tau=\chi g_v\mu_B B$. Here $M$ is computed using the so-called modern theory of orbital magnetization \cite{PhysRevB.109.L060409}:
\begin{align}
M &= \frac{e}{\hbar} \int \frac{d^2k}{(2\pi)^2}\sum_{n} n_F(\epsilon^{\lambda}_{n,\vec{k}}) \sum_{n'\neq n}\left(\epsilon^{\lambda}_{n,\vec{k}}+\epsilon^{\lambda}_{n',\vec{k}}-2\mu \right)\notag\\ 
&\frac{\operatorname{Im}\left(\bra{\psi_{n\vec{k}}^{\lambda}}\partial_x\hat{H}^{MF}_{\vec{k}}\ket{\psi_{n'\vec{k}}^{\lambda}}\bra{\psi_{n'\vec{k}}^{\lambda}}\partial_y\hat{H}^{MF}_{\vec{k}}\ket{\psi_{n\vec{k}}^{\lambda}}\right)}{(\epsilon^{\lambda}_{n,\vec{k}}-\epsilon^{\lambda}_{n',\vec{k}})^2}.
\end{align}
The linear and quadratic fitting shown in Fig.~\ref{fig:IVC-lagrange-multiplier}$(b)$ and $(c)$ yield the parameters for the phase-number action in RTG as $g_v \sim 3\;,\; \chi^{-1} \sim 0.1\text{meV}$. They are used to generate Fig.~\ref{fig:phase_OM_evolution}.



\textit{3-body Umklapp scattering:} At dilute doping where charge carriers are concentrated in the two opposite valleys, the Umklapp process that transfers electrons between these valleys necessitates at least a three-body scattering process. This is because a two-body scattering process does not provide sufficient momentum to match a reciprocal lattice vector. Such large-momentum transfer processes means a delta-function-like potential in the $K\cdot p$ continuum approximation. At second order in Coulomb potential and electron-phonon coupling, only one diagram contributes to this three-body scattering, as shown in Fig.~\ref{fig:Umklapp}. Under a lattice translation $\vec{a}$, $c_{\vec{K}} \rightarrow c_{\vec{K}} e^{i\vec{K} \cdot \vec{a}}$, this diagram picks up a phase $ e^{i6\vec{K} \cdot \vec{a}} =1 $ since $6\vec{K}$ is a reciprocal lattice vector.
\begin{figure}[t]
    \centering
\includegraphics[width=0.5\linewidth]{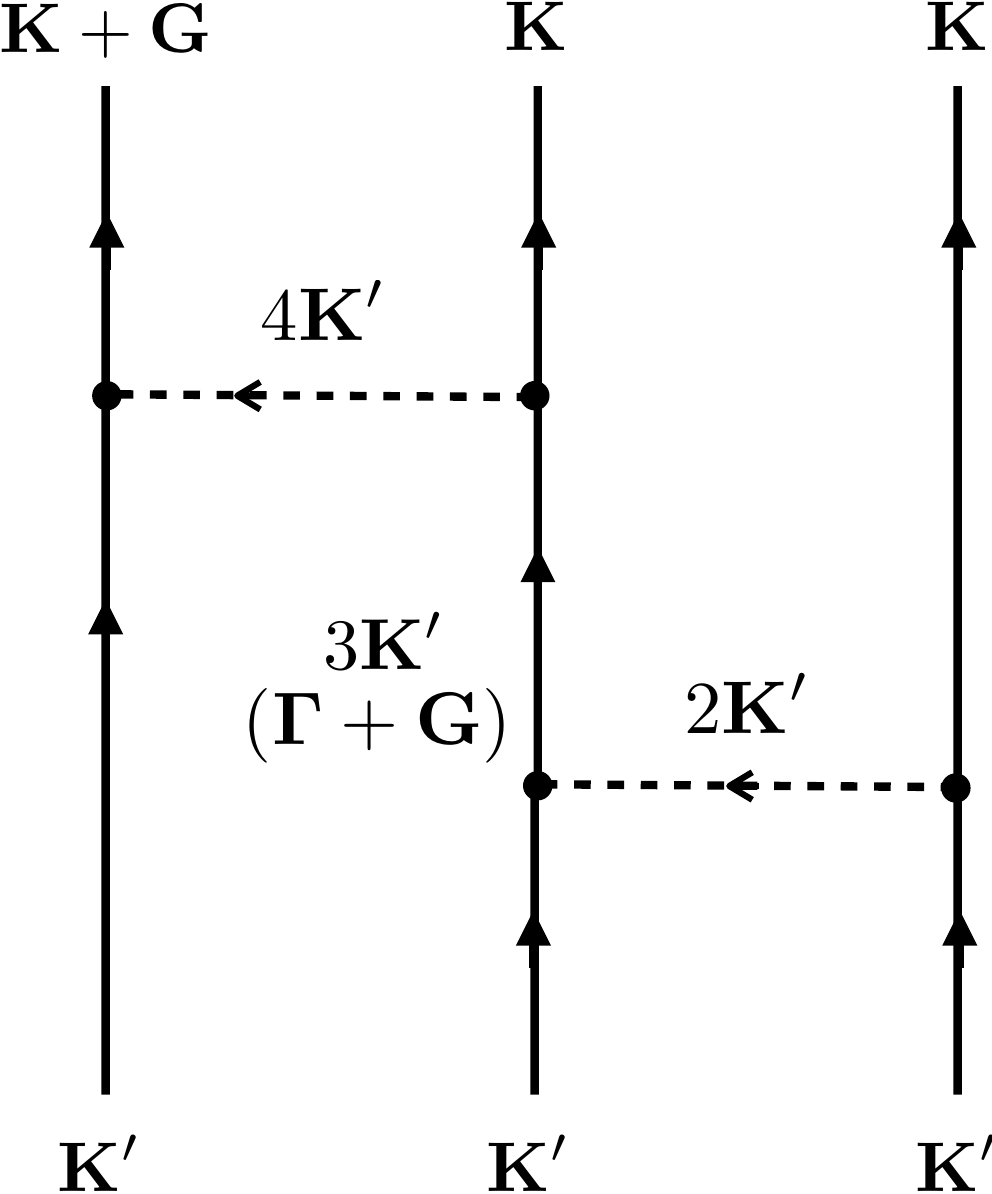}
    \caption{Feynman diagram of three-body Umklapp scattering process where three electrons from $K'$ valley are scattered to $K$ valley via an intermediate state at the $\Gamma$ point.}
    \label{fig:Umklapp}
\end{figure}
This scattering process begins with three electrons all initially in the same valley ($K'$). These electrons undergo two distinct scattering events to transition to the opposite valley $K$ modulo some reciprocal lattice vectors. First, two of the electrons scatter, moving one to the final valley 
$K$ and the other to an intermediate state at $3K$ (which corresponds to the 
$\Gamma$ point outside of the first Brillouin zone) with a matrix element 
$V(2\vec{K})$. Subsequently, the electron at the $\Gamma$ point scatters with the remaining electron at $K'$, transferring both to valley $K$ (modulo $G$) with matrix elements $V(4\vec{K})$. The strength of this scattering amplitude can therefore be estimated as follows:
\begin{equation}
    \gamma \sim \left| \frac{V(2\vec{K})V(4\vec{K})}{\epsilon_K - \epsilon_{\Gamma}} \right| \sim 0.5~ \mu\text{eV}
\end{equation}
In the last line, we used $V(2\vec{K}) \sim \frac{e^2}{\epsilon |2K|}n_e = 3~\text{meV}$ at $n_e=1\times10^{12}~\text{cm}^{-2}$ and \(|\epsilon_K - \epsilon_{\Gamma}| = 9~ \text{eV}\).

This type of lattice-scale interaction  includes higher order contributions from both electron-electron and electron-optical phonon interactions. We \cite{wei2025landau,xu2024influence,wolf2024magnetismdiluteelectrongas} and others \cite{PhysRevB.85.155439,knothe2020quartet,PhysRevLett.121.257001} have estimated the strength of two-body lattice-scale interactions, finding them significantly enhanced by the large density of states at graphene's 
$\pi$ band. In fact, without at least an order of magnitude enhancement in two-body lattice-scale interactions, their magnitudes would be insufficient compared to the Zeeman energy to explain the experimental findings \cite{zibrov2018even}. Fig.~8 in Ref.~\cite{xu2024influence} attempts to quantify this renormalization. Therefore, by treating $\gamma$ as a phenomenological constant and leveraging our better theoretical control of other constants in $F_{\text{IVC}}$, we propose detecting the oscillating magnetization—induced by a static magnetic field—as a practical experimental method to determine this parameter.

\textit{Discussions:} 
The universality class of $F_{\text{IVC}}$ is XY-like, and its topological defects are phase singularities known as merons \cite{moon1995spontaneous}. Below the Kosterlitz-Thouless (KT) transition temperature $T_{KT}$, or when $\gamma$ is sufficiently strong, merons and antimerons are bound, preventing phase slippage. However, for $T>T_{KT}$, unbound merons cause valley imbalance through phase slippage. This binding is crucial for the observability of our proposed phase-coherent transport phenomena. We have estimated $T_{KT}$ based on a microscopic calculation of IVC domain wall  \cite{supmat} and found $T_{KT}= (\pi/2) \rho\sim 3$K. This temperature is two orders of magnitude higher than the lowest temperatures accessible in many 2D transport experiments \cite{de_la_Barrera_2022,doi:10.1126/science.abm8386,seiler2022quantum,zhou_half_2021,arp2024intervalley,zhou_superconductivity_2021,lee2019gate,lee2014competition,liu2023spontaneous,Myhro_2018,wirth2022experimental,PhysRevB.79.125443,Han_2023,han2023orbital,morissette2025intertwinednematicitymultiferroicitynonlinear,PhysRevB.111.075103}, making our proposed phenomena experimentally feasible. 

The oscillating orbital magnetization in IVC can decay over time due to dissipative tunneling channels. In superconducting junctions, such dissipation arises from quasiparticle tunneling across the barrier, even at voltages below the superconducting gap. In our momentum-space analog of the AC Josephson effect, atomic-scale disorder can play a similar role by inducing random valley transfer, effectively introducing a "shunt resistance" between the $K$ and $K'$ valleys.
However, atomic-scale resolution STM studies \cite{nuckolls2023quantum,kim2023imaging,coissard2022imaging, liu2024visualizing}
have definitively identified extensive regions within graphene samples displaying IVC order, free from atomic-scale defects that could generate intervalley scattering.

In summary, we uncovered an interesting phase-coherent dynamics in IVC  and demonstrate their experimental observability by computing all parameters of the IVC phase-number free energy $F_{\text{IVC}}$ using a detailed microscopic theory. IVC has been theoretically identified in a wide range of two-dimensional materials. Notably, valley-projected moir\'e minibands with finite Chern numbers are common in graphene moiré superlattices \cite{xie2020nature,wagner2022global} and twisted homobilayer transition metal dichalcogenides (TMDs) \cite{reddy2023fractional}. Metallic IVC states have recently been proposed as the symmetry-broken phases that can mediate pairing glue for superconductivity observed in $5^\circ$ twisted WSe$_2$ \cite{fischer2024theory,munoz2025twist}. The phase-coherent transport discussed for the IVC metal in RTG can be directly generalized to this system, with modified parameters ($\chi^{-1}, g_v,\gamma$) readily computed using the methods we outlined above.
An interesting scenario arises in incompressible IVC states, which have been proposed to be the groundstate at integer fillings of moir\'e bands in twisted graphene \cite{wagner2022global,he2021competing,wu2021chern} multilayers and potentially also in twisted homobilayer MoTe$_2$ \cite{reddy2023fractional,sodemann2024halperin,zhang2024non}, which shows fractional quantum spin Hall effect \cite{kang2024evidence,kang2025time}. If an incompressible ground state has an IVC order and the moir\'e minibands in opposite valleys have opposite Chern numbers ($\pm C$), an external magnetic field induces valley imbalance through topological spectral flow: a magnetic flux $\Phi$ leads to exactly $C \Phi/\Phi_0$ occupied states per valley flowing adiabatically across the energy gap into unoccupied states where $\Phi_0=h/e$ is the flux quantum. States from valleys with opposite Chern numbers flow in opposite spectral directions, i.e.~states with $+|C|$ ($-|C|$) flow upward (downward) in energy. This directional spectral flow creates an valley-imbalance for the edge-states, generating finite orbital magnetization and a net edge current. Consequently, the distinction of our phase-coherent AC responses lies in the oscillating Hall current being delocalized throughout the bulk in IVC metal and being localized along the edges in IVC insulator.

Finally, we note that Ref.~\cite{wei2023weak} has shown that weak localization is enhanced in time-reversal-invariant IVC states. If this enhancement can be reliably distinguished from intervalley scattering effects -- also present in the normal metallic states -- it could provide a useful diagnostic for detecting IVC order. In contrast, the phase-coherent response we propose arises from interaction-driven three-body Umklapp scattering and is fundamentally distinct from disorder-related effects. As a result, our approach offers a more unambiguous route for identifying IVC.


\begin{acknowledgments}
\textit{Acknowledgments.}  
We acknowledge insightful discussions with J. Brill,  H. Fertig, G. Murthy, and N. Wei. We are grateful to the University of Kentucky Center for
Computational Sciences and Information Technology Services Research Computing for their support and use of the
Morgan Compute Cluster and associated research computing resources.
\end{acknowledgments}
\bibliography{references}
\newpage
\widetext
\begin{center}
\textbf{\large Supplementary Materials: Momentum-Space AC Josephson Effect and Intervalley Coherence in Multilayer Graphene}
\end{center}
\setcounter{equation}{0}
\setcounter{figure}{0}
\setcounter{table}{0}
\setcounter{page}{1}
\makeatletter
\renewcommand{\theequation}{S\arabic{equation}}
\renewcommand{\thefigure}{S\arabic{figure}}

\maketitle

\section{A: Rhombohedral Trilayer Graphene Bandstructure}
In this section of the supplementary material, we discuss the properties of the band Hamiltonian $\hat{H}_b$ using Slonczewski-Weiss-McClure (SWMC) parameters in rhombohedral trilayer graphene (RTG) used in the maintext. We discuss the effect of interlayer electric potential $(U)$ on the density of states (DOS).

\begin{figure*}[h]
    \centering   \includegraphics[width=0.8\linewidth]{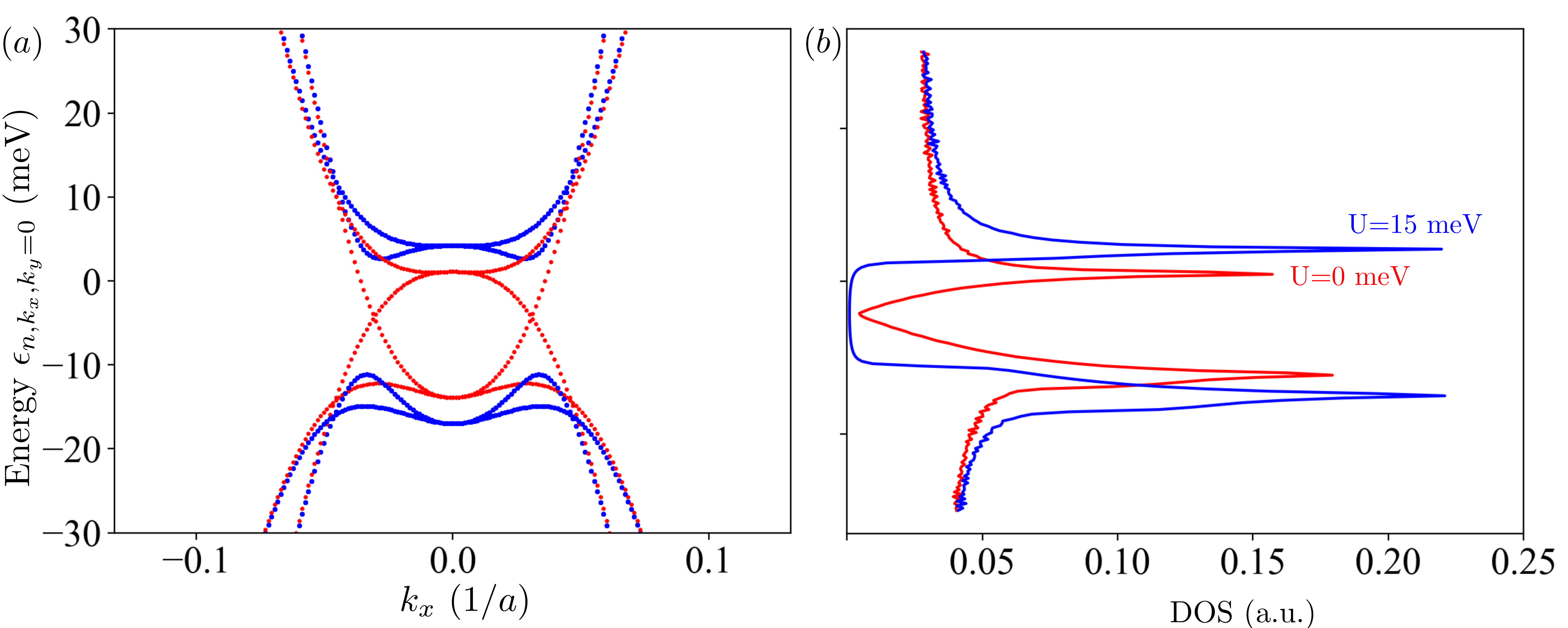}
    \caption{$(a)$ The non-interacting bandstructure of RTG at $U=0$ and $15$ meV vs $k_x$. $(b)$ The density of states (in arbitrary unit) enhances as $U$ increases near the band edge.}
    \label{fig:RTG_non-interacting}
\end{figure*}

RTG contains 3 layers of carbon atoms. The interlayer distance is $3.4$~{\AA} and each layer has two sublattices label by $\sigma=A,B$ which is separated by $1.42$ {\AA}, and the lattice constant of graphene is the distance between two nearest neighbor atoms from same sublattice, $a=2.46$~{\AA}. The layers of RTG are in a ABC stacking configuration.
Following Refs.~\cite{zhang_band_2010,koshino_trigonal_2009}, we consider a single electronic orbital ($\pi$) per atom, with spin $s=\pm 1/2$, and use the continuum model that can be derived from a tight-binding Hamiltonian by expanding the low-energy dispersion around each valley $\tau$ with $\tau=\pm 1$ at the corners of the Brillouin zone. We label the four flavor combinations of spin $s$ and valley $\tau$ with the flavor index $\alpha\equiv (\tau, s)$. In the basis $\ket{\psi_{\alpha}(\vec{k})}=(\psi_{\alpha A1}(\vec{k}), \psi_{\alpha B1}(\vec{k}), \psi_{\alpha A2}(\vec{k}), \psi_{\alpha B2}(\vec{k}), \psi_{\alpha A3}(\vec{k
}), \psi_{\alpha B3}(\vec{k}))$, the band Hamiltonian is given by
\begin{equation}
\hat{H}_b=\sum_{\substack{\vec{k}\\s=\pm1,\tau=\pm1}} \ket{\psi_{\tau,s}(\vec{k})} h_\tau(\vec{k}) \bra{\psi_{\tau,s}(\vec{k})}.
\end{equation}
The continuum Hamiltonian acting on the orbital space is different in opposite valley but spin-degenerate
\begin{align}\label{eq:continuum_hamiltonian}
    h_\tau(\vec{k}) = \begin{bmatrix}
    t(\vec{k}) + U_1 & t_{12}(\vec{k}) & t_{13} \\
    t_{12}^\dagger(\vec{k}) & t(\vec{k}) + U_2 & t_{12}(\vec{k}) \\
    t_{13}^\dagger & t_{12}^\dagger(\vec{k}) & t(\vec{k}) + U_3
    \end{bmatrix}_{6\times 6}.
\end{align}
Here $\vec{k}=(k_x,k_y)$ is the Bloch momentum measured with respect to the center of valley $\tau$. This Hamiltonian contains matrices with intra-layer hopping amplitudes ($t$), nearest-layer hopping amplitudes ($t_{12}$), and next-nearest-layer hopping amplitudes ($t_{13}$), as well as possible potential differences between different layers and/or sublattices due to external gates and/or broken symmetries (terms $U_i$). Explicitly, the various hopping amplitudes are given by
\begin{align}
    t(\vec{k}) = \begin{bmatrix}
      0 & v_0 \pi^\dagger \\
      v_0 \pi & 0
    \end{bmatrix}, \quad 
    t_{12}(\vec{k}) = \begin{bmatrix}
      -v_4 \pi^\dagger & v_3 \pi \\
      \gamma_1 & -v_4 \pi^\dagger
    \end{bmatrix}, \quad
    t_{13} = \begin{bmatrix}
      0 & \gamma_2/2 \\
      0 & 0
    \end{bmatrix}
\end{align}
where $\pi = \tau k_x+i k_y$ is a linear momentum, and $\gamma_i$ ($i=0,\dots, 4$) are hopping amplitudes of the tight-binding model with corresponding velocity parameters $v_i=(\sqrt{3}/2)a\gamma_i/\hbar$. The ${U_i}$ terms are
\begin{align}
    U_1 = \begin{bmatrix}
      \frac{\Delta+\delta+U}{2} & 0 \\
      0 & \frac{\Delta}{2}
    \end{bmatrix}, \quad 
    U_2 = \begin{bmatrix}
      -\Delta & 0 \\
      0 & -\Delta
    \end{bmatrix}, \quad
    U_3 = \begin{bmatrix}
      \frac{\Delta}{2} & 0 \\
      0 & \frac{\Delta+\delta-U}{2}
    \end{bmatrix}
    \label{eq:Ui}
\end{align}

We use the model parameters listed in Table I, which are chosen to match quantum oscillation frequency signatures in Ref.~\cite{zhou_half_2021}. The tunable parameter $U$ in Eq.~\ref{eq:Ui}  is the interlayer potential, and is proportional to out-of-plane electric displacement field. Fig.~\ref{fig:RTG_non-interacting}.$(a)$ shows the bandstructure along $x$ direction at $k_y=0$ for $U=0$ (in red) and $U=15$ meV (in blue). At $U=0$, the valance and conduction bands form Dirac touching near the high symmetric $K,K'$ points. At finite $U$, the Dirac touching is lifted and the band edges become flatter. Consequently, DOS increases near the band edge with increasing $U$, as shown in Fig.~\ref{fig:RTG_non-interacting}.(b).

\begin{table}[t]
\caption{\label{tab:graphene_params_ABC}Tight-binding parameters (in eV) for rhombohedral trilayer graphene, see also Refs.~\cite{zhang_band_2010,koshino_trigonal_2009,zhou_half_2021}.}
\begin{ruledtabular}
\begin{tabular}{llllllll}
$\gamma_0$ & $\gamma_1$ & $\gamma_2$ & $\gamma_3$ & $\gamma_4$ & $U$ & $\Delta$ & $\delta$ \\
$3.160$ & $0.380$ & $-0.015$ & $-0.290$ & $0.141$ & $0.030$ & $-0.0023$ & $-0.0105$ \\
\end{tabular} 
\end{ruledtabular} 
\end{table}

\section{B: Intervalley-coherent state}
\begin{figure*}[h]
    \centering   \includegraphics[width=0.7\linewidth]{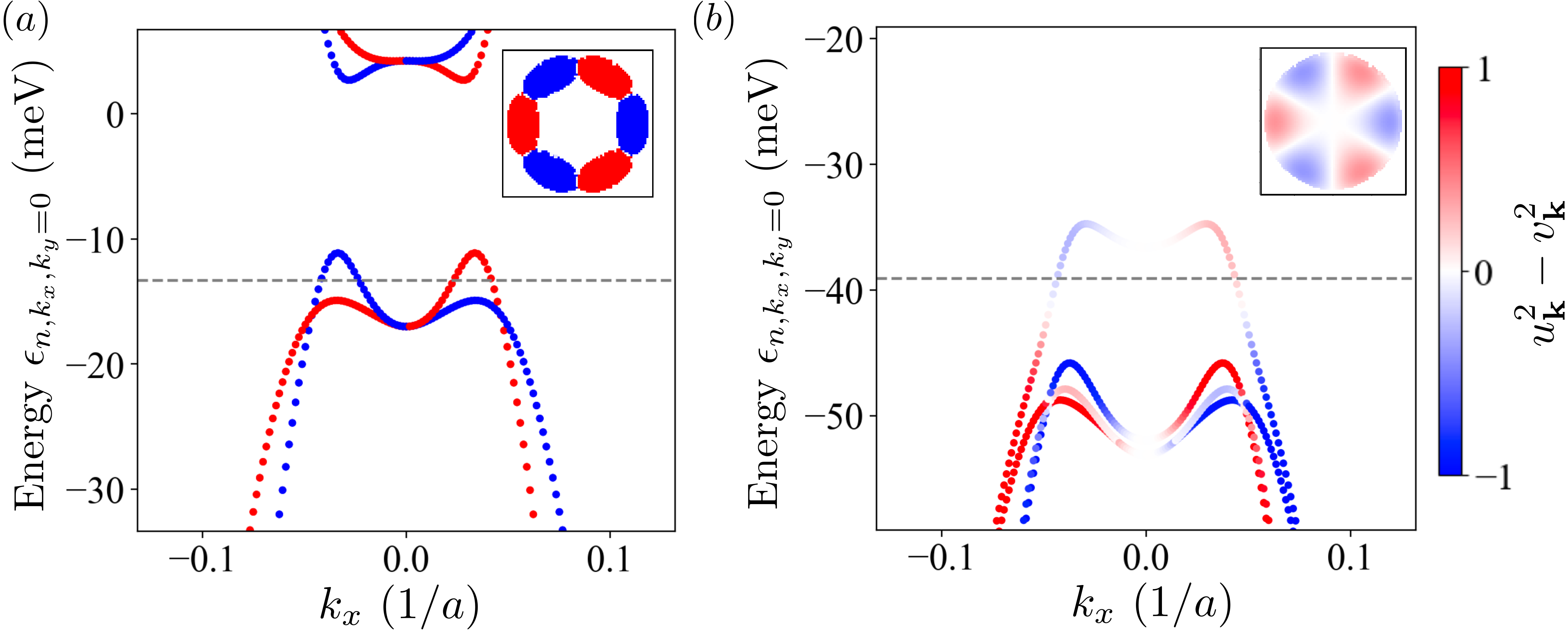}
    \caption{$(a)$  RTG non-interacting bandstructure at $U=15$ meV along $x$-direction at $k_y=0$. The red and blue color indicating the bands coming from $K$ and $K'$ valleys. The dashed line marks the Fermi level, and the inset indicating the Fermi surface at $(n_e, U)=-2.7\times 10^{11} \text{cm}^{-2},15$ meV. $(b)$ Coulomb interaction spontaneously mixes the valleys, and forms spin-polarized IVC groundstate. The color bar indicates cosine of the valley mixing angle. The inset shows the distribution of valley mixing over the Fermi surface.}
    \label{fig:IVC}
\end{figure*}

In this subsection, we provide further details on the inter-valley coherent (IVC) state in rhombohedral trilayer graphene (RTG). In Fig.~\ref{fig:IVC}(a), we plot the non-interacting band dispersion of the SWMC model, $\epsilon_K(k_x, k_y = 0)$ and $\epsilon_{K'}(k_x, k_y = 0)$, as a function of $k_x$. The bands from the two valleys are not hybridized and are color-coded red and blue for the $K$ and $K'$ valleys, respectively. Each valley hosts three small hole-like Fermi surfaces that respect the lattice’s $C_3$ rotational symmetry. The Fermi surface configuration in the opposite valley is rotated by 180 degrees due to time-reversal symmetry.

Figure~\ref{fig:IVC}(b) shows the bandstructure after a self-consistent Hartree-Fock calculation at fixed carrier density $n_e$. Instead of three tiny Fermi surfaces per valley (six in total including spin), the mean-field solution results a single spin-polarized Fermi surface with a momentum-dependent valley-mixing angle. The physical property of this state remains invariant under changes to the global relative phase $e^{i\phi}$ between the two valleys. This defines the inter-valley coherent (IVC) state.

\section{C: Intervalley-coherent state domain wall}
\begin{figure*}[h]
    \centering   \includegraphics[width=0.8\linewidth]{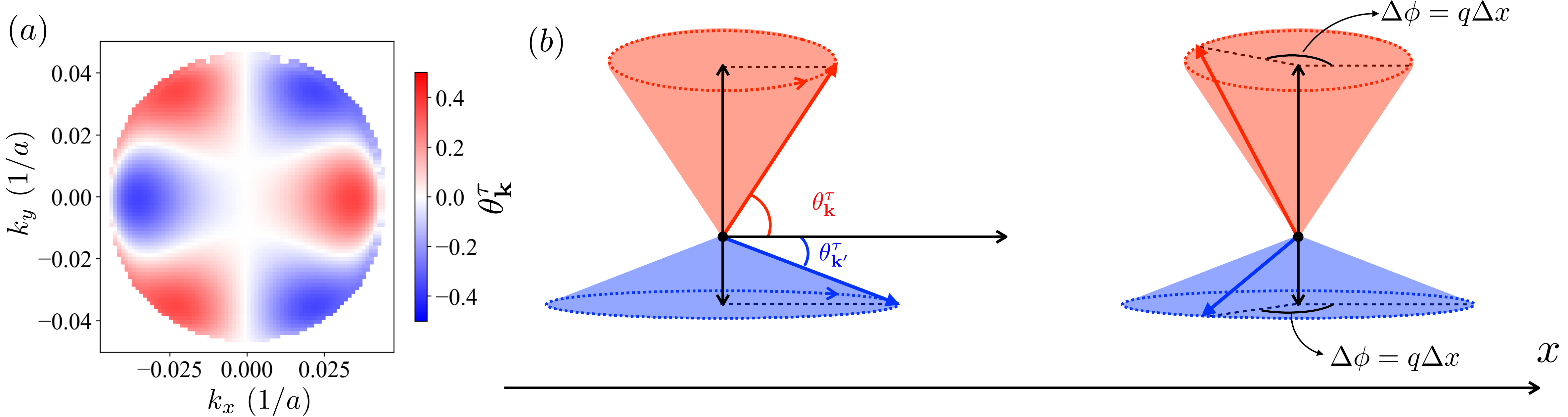}
    \caption{$(a)$  Variation of valley-pseudospin cone-angle for a converged IVC domain wall solution  $\vec{q}=0.05~\vec{e}_x/a$.  $(b)$  Schematic diagram represents the valley pseudospins (shown by the red and blue arrows) of different momenta precess with different cone angle as it moves along the $x$-axis.}
    \label{fig:IVC-domain-wall}
\end{figure*}

In the appendix, we discuss the microscopic calculation to determine the energy cost of creating an IVC domain wall and to extract the valley-stiffness parameter to estimate the Kosterlitz-Thouless transition. In this section, we use the valley pseudospin language instead of the exciton language.  We specifically solve for a Slater-determinant where the azimuthal-angle of all the Bloch states $\phi(r)$ increases linearly in space along a certain direction $\phi(\vec{r})=\vec{q}\cdot \vec{r}$, causing the Bloch states to precess as they move through the material. In these calculations, we set $\vec{r}=x \hat{e}_x$ direction, ignoring any small $C_3$ anisotropic dependence.

Then, following the argument of  Ref.~\cite{herring1952energy}, we shift the band Hamiltonian in opposite valley in opposite direction by $q_x/2$:
\begin{align}
   \hat{H}_b(\vec{q})=\sum_{k,\tau,s,\sigma}\; h_{\sigma,\sigma'}(\tau (k_x+q_x/2),k_y) \;c^\dagger_{\tau,\sigma,s}(k_x+\tau q_x/2,k_y)  c_{\tau,\sigma',s} (k_x+\tau q/2,k_y),
   \label{eq:valley-shifted-Hamiltonian}
\end{align}
where $h_{\sigma,\sigma'}(\vec{k})$ is the matrix element of $h_{\tau}(\vec{k})$ described by Eq.~\ref{eq:continuum_hamiltonian} in the sublattice space. This shift changes the Hamiltonian's symmetry, reducing it from independent continuous translation and valley 
$U(1)$ rotation symmetries to a combined translation plus valley 
$U(1)$ rotation rotation symmetry. This is identical to the symmetry of valley-density wave. At small $q$,  it is sufficient to consider only two reciprocal lattice vectors, $\vec{G}=\pm q/2 ~\vec{e}_x$. Importantly, by excluding the $G=0$ component in the Hamiltonian, we prevent any uniform valley polarization, so that the converged solution represents a unidirectional domain wall without net valley polarization.

We then incorporate the density-density interaction term $V$ from Eq.~\ref{eq:Coulomb-pot} in the main text into the shifted Hamiltonian, and solve the resulting Hamiltonian, $\hat{H}(\vec{q})=\hat{H}_b(\vec{q})+\hat{V}$, within the mean-field approximation for various values of $\vec{q}$. In this approximation, the full interacting Hamiltonian is replaced by a single-particle-like mean-field Hamiltonian, $\hat{H}_{MF}(\vec{q})$, whose matrix elements are determined self-consistently. Solving the mean-field equations iteratively yields the complete eigenspectrum $\epsilon_{n\vec{k}}^{\vec{q}}$ and $|\psi_{n\vec{k}}^{\vec{q}}\rangle$ from which we construct the converged density matrix $\hat{\rho}(\vec{q})=\sum_{n,\vec{k}} n_F(\epsilon_{n\vec{k}}^{\vec{q}})|\psi_{n\vec{k}}^{\vec{q}}\rangle \langle \psi_{n\vec{k}}^{\vec{q}}|\equiv \sum_{\vec{k}} \hat{\rho}_{\vec{k}}(\vec{q})$. From this density matrix, we compute the total energy
\begin{equation}
E(\vec{q})=\frac{1}{2}\text{Tr}\left[(\hat{H}_b(\vec{q})+\hat{H}_{MF}(\vec{q}))\hat{\rho}(\vec{q})\right]
\end{equation}
and the valley-pseudospin cone-angle $\theta^\tau_{\vec{k}}(\vec{q})$ defined by
\begin{equation}
\tan{\theta^\tau_{\vec{k}}(\vec{q})}=\frac{\operatorname{Tr}\left[\hat{\rho}_{\vec{k}}(\vec{q})\hat{\tau}_z\right]}{\operatorname{Tr}\left[\hat{\rho}_{\vec{k}}(\vec{q})\hat{\tau}_x\right]},
\end{equation}
where $\hat{\tau}_{x,z}$ are Pauli matrices acting on the valley degrees of freedom. In exciton language, the cone angle $\theta^\tau_{\vec{k}}(\vec{q})$ corresponds to the relative weight between the no-exciton and single-exciton states, $\cos(\theta^\tau_{\vec{k}}(\vec{q}))=u^2_{\vec{k}}(\vec{q})-v_{\vec{k}}^2(\vec{q})$ where $u$ and $v$ are defined in Eq.~1 of the maintext.

Fig.~\ref{fig:IVC-domain-wall}(a) illustrates the variation of cone-angle  inside the Fermi sea. We note that while different Bloch states exhibit different cone angles, they all precess with the same spatial periodicity since $\phi=q_xx$ is independent of the Bloch state's momentum $\vec{k}$. Consequently, these states precess at different angular velocities as they move through the material. This picture is illustrated in Fig.~\ref{fig:IVC-domain-wall}(b) for two Bloch states with different momentum $\vec{k}$ and $\vec{k}'$.

\begin{figure*}[h]
    \centering   \includegraphics[width=0.35\linewidth]{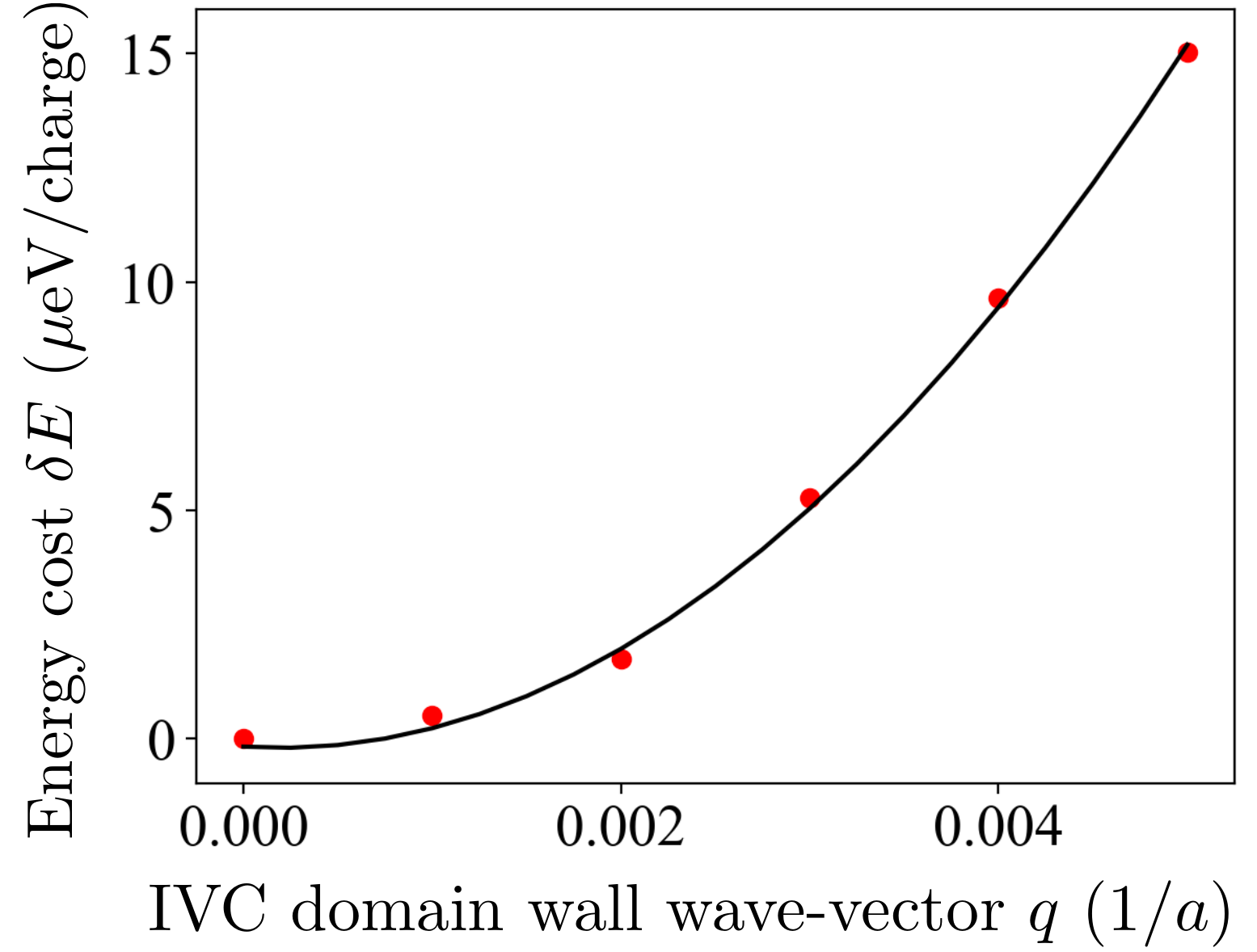}
    \caption{The energy cost per charge to create domain wall of wave-vector $|\vec{q}|$ shows quadratic behavior, and is proportional to valley-stiffness constant. This energy cost is calculated at $(n_e, U)=-2.7\times 10^{11} \text{cm}^{-2},15$ meV.}
    \label{fig:IVC-domain-wall-enegy}
\end{figure*}
In Fig.~\ref{fig:IVC-domain-wall-enegy}, we plot the energy cost per charge of IVC domain wall $\delta E(q)=(E(q)-E(0))/|N_e|$ measured with respect to the uniform state, as function of wave vector magnitude $ q$ for parameters $(n_e, U)=(-2.7\times 10^{11}, \text{cm}^{-2}, 15, \text{meV})$. Here, $N_e=n_eA$ is the total charge in the sample with $n_e$ being the charge density and $A$ being the sample area.  The valley-stiffness constant $\rho$ is defined as the quadratic expansion of the energy change per charge in terms of $q$:
\begin{align}
 \delta E(q) =\frac{\rho}{|n_e|}\cdot  q^2 + O(q^4).
\end{align}
 This calculation yields an estimated valley-stiffness constant $\rho \sim 0.16 ~\text{meV}$.

\end{document}